\documentclass[prb,showpacs,superscriptaddress,twocolumn,notitlepage]{revtex4-1}
\usepackage{graphicx, subfigure}

\usepackage{amsmath}
\usepackage{amsfonts}
\usepackage{graphicx,subfigure}

\usepackage{color}
\usepackage[usenames,dvipsnames]{xcolor}

\usepackage{natbib}
\usepackage[final=true]{hyperref}

\begin{document}
\title{Molecular dynamics-based refinement of nanodiamond size measurements obtained with dynamic light scattering}

\author{S.~V.~Koniakhin}
\email{kon@mail.ioffe.ru}
\affiliation{St. Petersburg Academic University, Khlopina 8/3, 194021 St. Petersburg, Russia}
\affiliation{Ioffe Physical-Technical Institute of the Russian Academy of Sciences, 194021 St.~Petersburg, Russia}

\author{I.~E.~Eliseev}
\affiliation{St. Petersburg Academic University, Khlopina 8/3, 194021 St. Petersburg, Russia}

\author{I.~N.~Terterov}
\affiliation{St. Petersburg Academic University, Khlopina 8/3, 194021 St. Petersburg, Russia}

\author{A.~V.~Shvidchenko}
\affiliation{Ioffe Physical-Technical Institute of the Russian Academy of Sciences, 194021 St.~Petersburg, Russia}

\author{E.~D.~Eidelman}
\affiliation{Ioffe Physical-Technical Institute of the Russian Academy of Sciences, 194021 St.~Petersburg, Russia}

\author{M.~V.~Dubina}
\affiliation{St. Petersburg Academic University, Khlopina 8/3, 194021 St. Petersburg, Russia}

\begin{abstract}
The determination of particle size by dynamic light scattering uses the Stokes-Einstein relation, which can break down for nanoscale objects. Here we employ a molecular dynamics simulation of fully solvated 1-5 nm carbon nanoparticles for the refinement of the experimental data obtained for nanodiamonds in water by using dynamic light scattering. We performed molecular dynamics simulations in differently sized boxes and calculated nanoparticles diffusion coefficients using the velocity autocorrelation function and mean-square displacement. We found that the predictions of the Stokes-Einstein relation are accurate for nanoparticles larger than 3\,nm while for smaller nanoparticles the diffusion coefficient should be corrected and different boundary conditions should be taken into account.

%\keywords{dynamic light scattering, molecular dynamics, diffusion, Stokes-Einstein relation, nanodiamonds}

\end{abstract}

\maketitle   % please do not remove

\section{Introduction}
To interpret the data obtained in a dynamic laser light scattering experiment, the Stokes-Einstein relation must be met, which requires correct use of the macroscopic viscosity of liquid and of the relations of classical hydrodynamics. One must define the exact lower limit on the particle size for which this relation is applicable \citep{PhysRevE.80.061204,TutejaEtAl2007,ould-kaddour:154514,BounadaryCond}.

Dynamic light scattering (DLS) is a powerful nondestructive method that is employed in physics, chemistry, biology, and nanotechnology to determine the size of objects suspended in a solvent. Although dynamic light scattering is considered to be a technique developed for particle size measurement, the immediate result from a DLS experiment is the autocorrelation function of the scattered light intensity. For a suspension that contains particles of various sizes, the autocorrelation function takes the following form \citep{johnson1994lasers,DLS_JNR}:
\begin{equation} \label{DLS}
G(\tau)=C_0+\left( \sum{C_i \exp (-\Gamma_i \tau)}\right)^2,
\end{equation}
where the intensity correlation decay rate $\Gamma_i$ that corresponds to the i-th type of particles. $\Gamma_i$ is expressed in terms of diffusion coefficient $D_i$, wavelength $\lambda$ of the light undergoing scattering, solvent refraction coefficient $n$ and scattering angle $\theta$ as
\begin{equation} \label{Gamma_i}
\Gamma_i=D_i \left( \frac{4 \pi n}{\lambda} \sin{(\theta / 2)} \right)^2.
\end{equation}
Each parameter $C_i$ is defined by the relation between the fractions of the particles of the corresponding size in the suspension and the intensities of the light scattered by them. The diffusion coefficient $D$ and the size are coupled through the Stokes-Einstein relation
\begin{equation} \label{Diff_Rad}
D=\frac{k_BT}{6\pi\eta R},
\end{equation}
where $T$ is the temperature of the system, $\eta$ is the viscosity of the liquid, $R$ is the particle radius, and $k_B$ is the Boltzmann constant. The viscosity of water at room temperature $8.9\cdot 10^{-4}\,\text{Pa}\cdot\text{s}$ was used in all calculations.

The type of hydrodynamic boundary conditions (stick or slip) on the surface of the particle affects the coefficient in the Stokes-Einstein relation. The most common form of the Stokes-Einstein relation \eqref{Diff_Rad} corresponds to the stick boundary conditions. It is considered to describe the case of macroscopic particles. For slip boundary conditions one writes 4 instead of 6 in the denominator of relation \eqref{Diff_Rad}. In ref. \citep{PhysRevLett.101.226101} the authors show that the simulated slippage length for water and various surfaces (including the diamond-like surfaces) can reach dozens of nanometers, which exceeds the characteristic size of nanodiamonds and makes discussable the type of the actual hydrodynamic boundary conditions.

The efficacy of applying molecular dynamics to the derivation of diffusion characteristics of particles in the solid ball model has been demonstrated previously \citep{ould-kaddour:154514,Rudyak}. The appearance of a solvate shell that coats the surface of a nanoparticle may prove to have interesting side effects \citep{C2NR33512C}. The application of simulations in the context of fully atomistic molecular dynamics would permit microscopic features of the interaction between the medium and the particles to be taken into consideration.

This work accounts for the atomic structures of the particles and the solvent, which are aspects that should become essential for determination of particle sizes on the nanometer scale. The aim of this study is to check applicability of molecular dynamics for estimation of the particles diffusion properties in the context of global problems of computational methods for estimation of thermodynamic quantities \citep{DarkSide}. In particular we apply molecular dynamics simulation to determine the type of hydrodynamic boundary conditions for differently sized nanodiamonds.

\section{Experiment}

A particular feature of the structure of detonation nanodiamonds is their characteristic monocrystallite size of 4\,nm \citep{0022-3727-40-20-S14}, which is energetically favorable under the conditions of the growth process. This value has been obtained using X-ray scattering and TEM data \citep{Ozerin}, processing of the Raman scattering spectra in the phonon confinement model \citep{Raman_FTT_Pevtsova} and computer simulation \citep{Raty}.

The technique employed for the preparation of this sample has been described previously \citep{DLS_Shvid,Aleksenskiy}. All suspensions of detonation nanodiamonds contain both single crystallites and their aggregates of size up to 100\,nm. Light scattering intensity of a single particle or aggregate is proportional to the square of its volume. It is the reason why the optical signal from single particles is always much more weak than the signal from aggregates. The bimodal model assuming that suspension contains particles of two characteristic sizes ($i$ = 1, 2) allows to extrude a small contribution to its optical properties from single 4\,nm nanoparticles only. For instance it has been found to be appropriate for the description of the optical density of such suspensions \citep{Labeling_my}.

\begin{figure}
\includegraphics[width=1.0\linewidth]{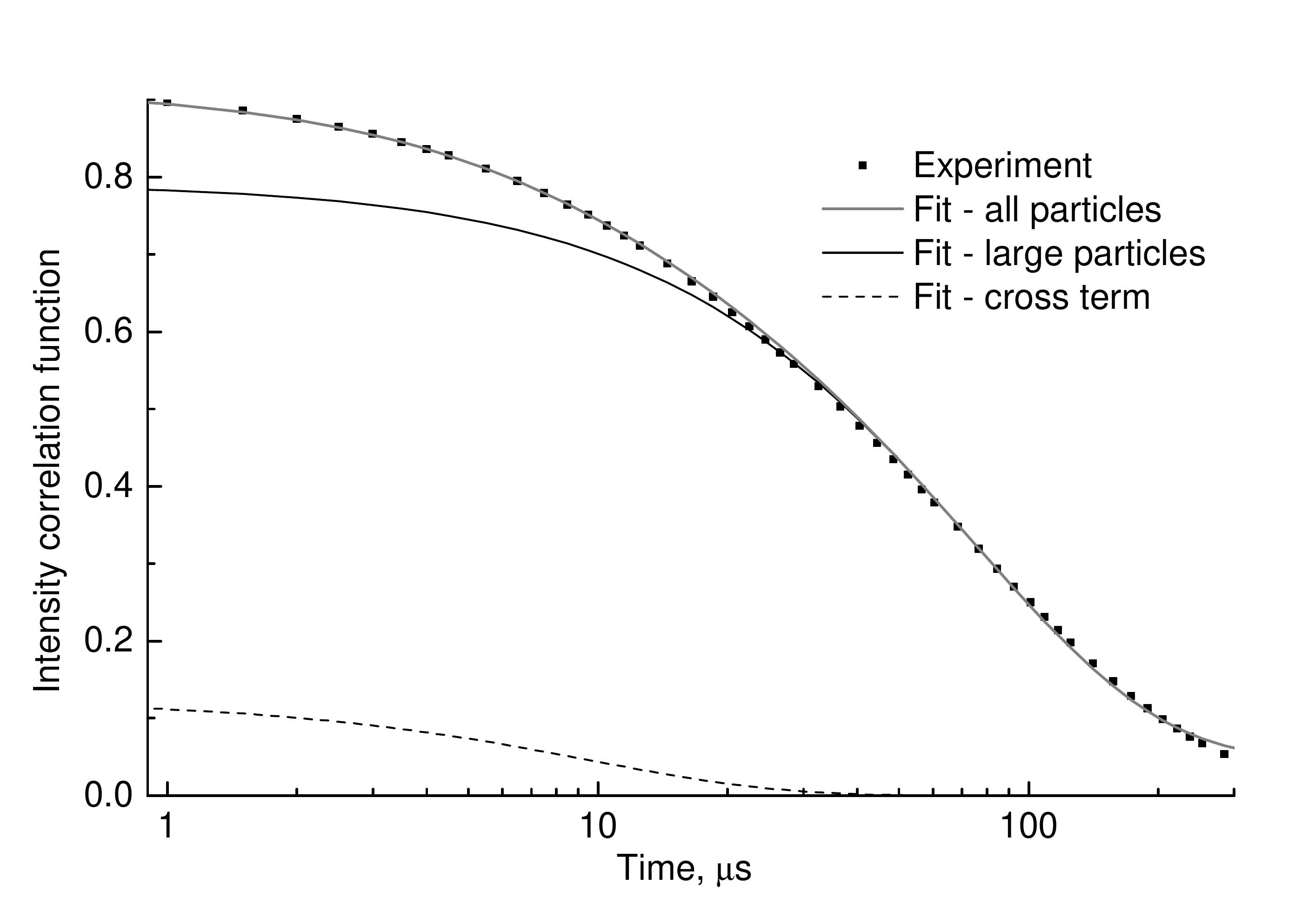}
\caption{Autocorrelation function of the scattering intensity for suspension of disaggregated nanodiamonds (squares) measured by dynamic laser light scattering and the result of its approximation with relation \eqref{DLS} (grey line). The solid black curve represents the contribution from large particles only, and the dashed black curve represents the contribution of the cross-term.}
\label{DLS}
\end{figure}

Figure \ref{DLS} displays the correlation function obtained with the Malvern Zetasizer instrument ($\theta = 173^{\circ}$, $\lambda=633$\,nm) for suspension of disaggregated nanodiamonds. The correlation function was approximated manually using relation \eqref{DLS}, where $i=1,2$. The data corresponding to long times were first used to determine parameter $C_0\approx 0.047$ and parameters $C_2\approx 0.86$ and $\Gamma_2\approx (152 \, \mu \text{s})^{-1}$ for large particles, after which the difference between the experimental data and the contribution from large particles in the region corresponding to short times was attributed to the contribution from small particles and used to approximate the remaining parameters: $C_1\approx 0.07$ and $\Gamma_1\approx (10.3\,\mu \text{s})^{-1}$. It can be seen that cross term $2C_1 C_2 \exp\left(-(\Gamma_1+\Gamma_2)\tau \right)$ is dominant for short times because $C_1 \ll C_2$.

The diffusion coefficient of single particles calculated with relation \eqref{Gamma_i} is $D_1=1.1 \times 10^{-6} \, \text{cm}^2\text{s}^{-1}$. If one knows independently the estimate of the nanoparticles size (4\,nm for detonation nanodiamonds), the dynamic light scattering method of measuring their diffusion coefficient becomes an experimental way to verify the Stokes-Einstein relation.

\section{Simulation details}

For comparison with analytical expression \eqref{Diff_Rad} and the DLS data, the diffusion coefficients of nanoparticles were calculated using molecular dynamics (MD) in an explicit solvent in the Gromacs 4.5.4 package\citep{doi:10.1021/ct700301q}. The nanoparticles were modeled as carbon nanocrystals with a diamond-type lattice (sp$^3$ hybridization) of an approximately spherical shape with diameters of 1\,nm, 1.5\,nm, 2\,nm, 3\,nm, 4\,nm, and 5\,nm. The surface atoms of nanoparticle lacking some neighbors were considered as bonded with the implicit hydrogens. This model describes idealized hydrophobic objects that nonetheless have an atomic structure.

The carbon-carbon van der Waals interactions were computed using Lennard-Jones potential with the parameters proposed in DREIDING forcefield \citep{doi:10.1021/j100389a010}. This forcefield also includes the parameters for the hydrated carbon atoms (see table VII from \citep{doi:10.1021/j100389a010}).

The solvent was modeled using TIP4P/2005 water model \citep{water} which provides accurate estimates of water viscosity compared to other popular water models (SPC/E, TIP4P and TIP4P/Ew). The latter was shown using the molecular dynamics simulations of self-diffusion coefficient and stress tensor in bulk water \citep{0953-8984-24-28-284117} and water flow between two surfaces \citep{viscosity}. The water-carbon interactions were calculated using common mixing rules $\varepsilon_{ij}=\sqrt{\varepsilon_{ii}\varepsilon_{jj}}, \sigma_{ij}=\sqrt{\sigma_{ii}\sigma_{jj}}$.

For each size of the nanoparticle 5 systems with different cubic box size ranging from 3 to 16\,nm were prepared. Two MD trajectories of 10\,ns and 50\,ns lengths with 2\,fs time step were calculated for each system. The shorter trajectories with 0.05\,ps velocity sampling interval was used for the velocity autocorrelation function (VAF) analysis while the longer trajectory sampled every 1\,ps was used for the mean-square displacement (MSD) calculation. Periodic boundary conditions were used in all simulations. The solvent molecules were maintained at 298\,K by the Berendsen thermostat.

The results of the simulations were used to calculate the diffusion coefficient in two ways. The first method involved the application of the Kubo relation \citep{JPSJ.12.570} coupling the diffusion coefficient to the VAF of particles

$C_{v} (\tau) =\langle \textbf{v} (t) \textbf{v} (t+ \tau) \rangle_{t}$:

\begin{equation} \label{kubo1}
D=\frac{1}{3}\int_0^{\infty} \langle \textbf{v} (t) \textbf{v} (t+ \tau) \rangle_{t} d \tau.
\end{equation}

The second method treats the determination of the diffusion coefficient in terms of the MSD of a random-walking object from its initial position:

\begin{equation} \label{Sqr}
\langle \textbf{r}^2 \rangle=6Dt.
\end{equation}

The angle brackets here represent an ensemble average that is computed over trajectories calculated using the molecular dynamics method.

\section{Results and discussions}

\begin{figure}
\includegraphics[width=1.0\linewidth]{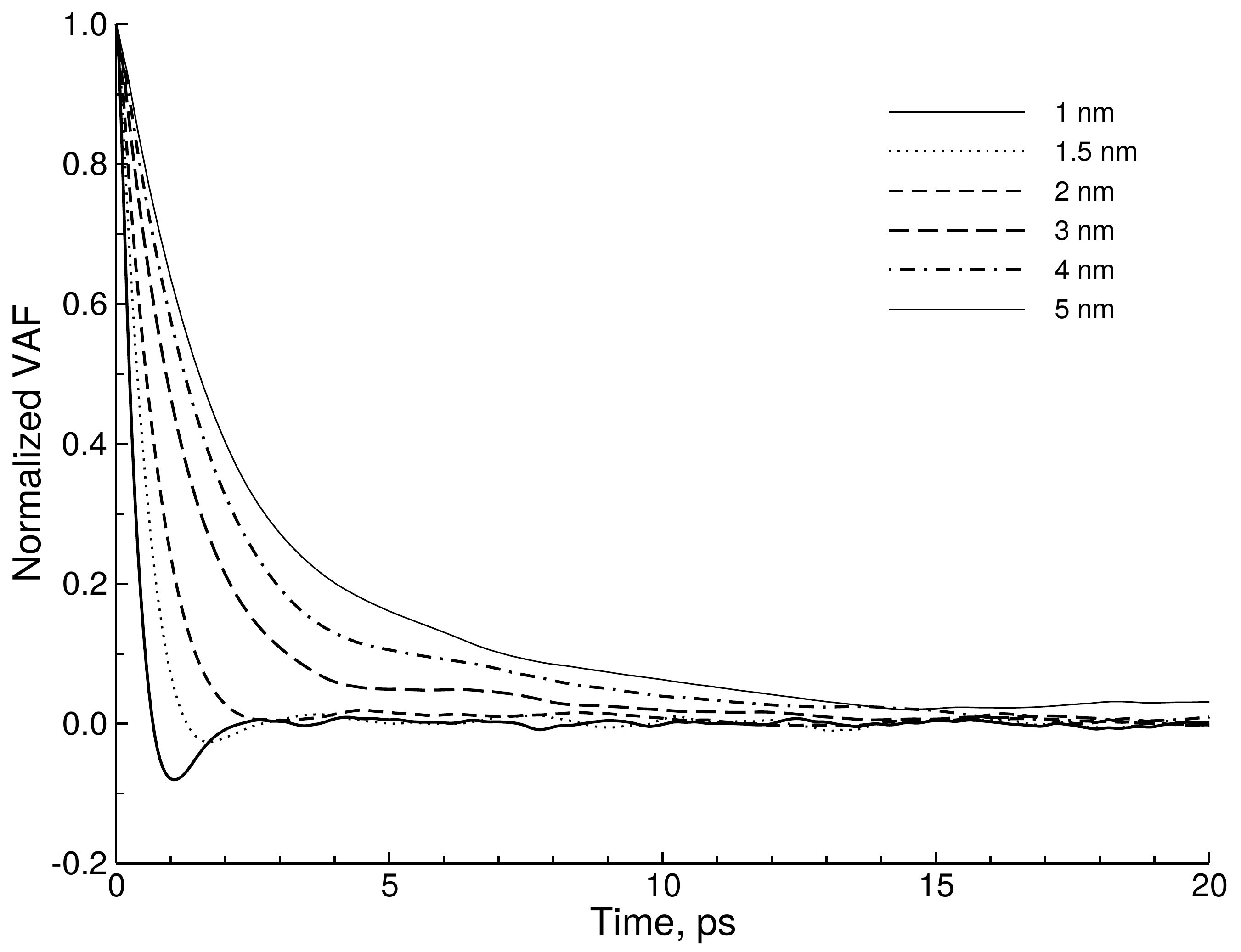}
\caption{The averaged over 100 trajectories normalized velocity autocorrelation functions for particles of different sizes.}
\label{fig_vaf}
\end{figure}

For determining the nanoparticles diffusion coefficient using the VAF approach the whole 10\,ns trajectory was considered as 100 independent trajectories of 100\,ps, which allowed to calculate the mean diffusion coefficient and its standard deviation. Fig. 2 shows time dependencies of the normalized to unity velocity autocorrelation functions for particles of various sizes for the largest of simulated boxes. One sees that for the solute size of about 3\,nm the velocity autocorrelation function resembling a damped oscillating motion transforms to a strictly decaying function. Integral \eqref{kubo1} is taken numerically in the range from 0 to 25\,ps.

\begin{figure}
\includegraphics[width=1.0\linewidth]{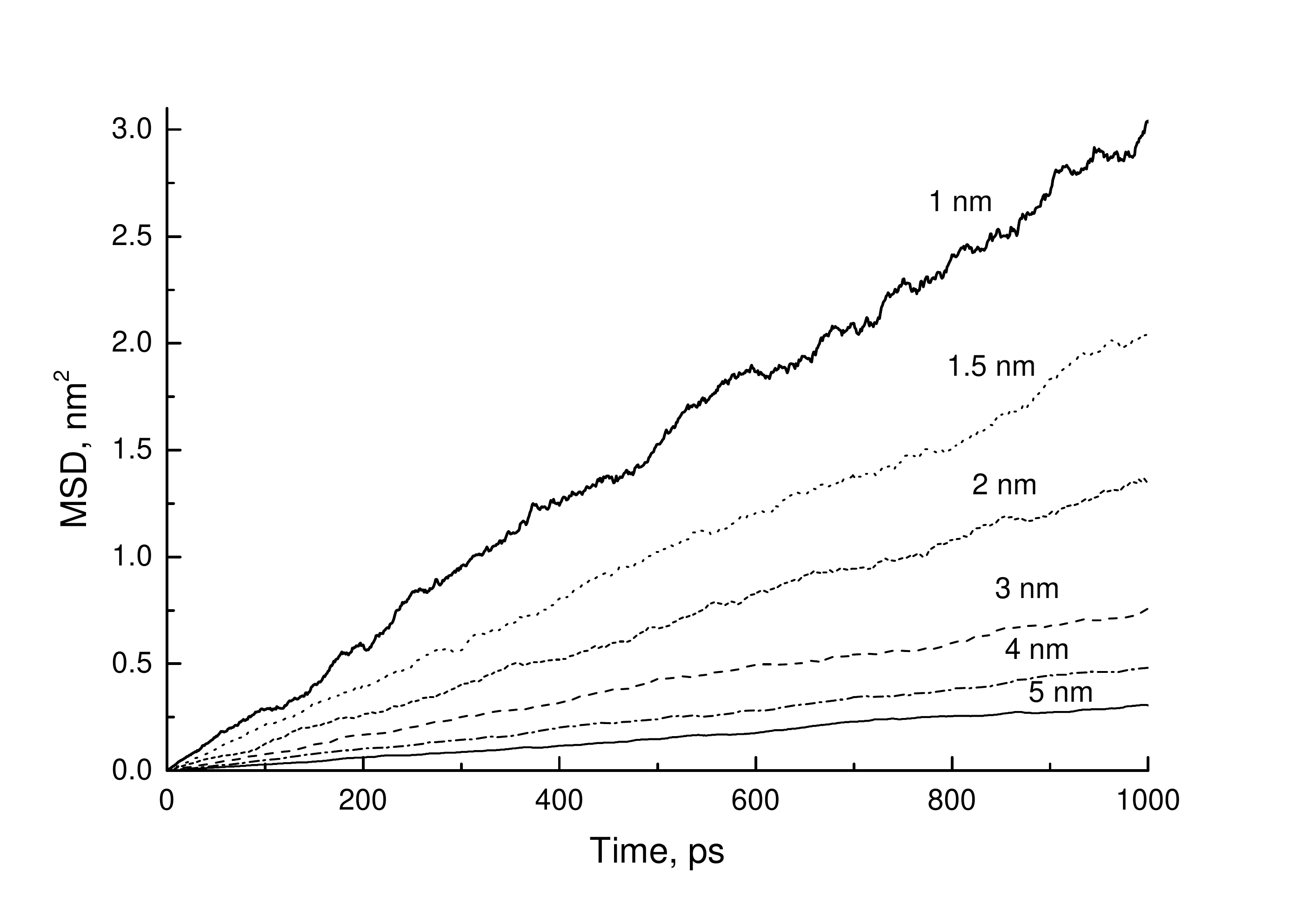}
\caption{Mean-square displacement for differently sized particles vs. time for largest simulation boxes (12-16\,nm).}
\label{fig_msd}
\end{figure}

The trajectory length related to the MSD approach was $t = 50$\,ns. The whole simulated trajectory was considered as $t/t_R$ trajectories of successively decreasing lengths: $t$, $t-t_R$ etc. The averaging of the MSD was provided over 100 trajectories. Restart time $t_R$ was chosen to be 500\,ps, however the obtained results exhibited weak dependencies of diffusion coefficients on fitting range and $t_R$. The dependence of the mean-square displacement on time is shown in fig. \ref{fig_msd} for particles of all sizes. The error of diffusion coefficient was estimated as a standard deviation of mean values of the MSD calculated for 10 groups of 10 trajectories.

Previously it was shown \citep{PolymerChain,doi:10.1021/jp0477147} that the diffusion coefficient $D_{PBC}(L)$ of nanoparticle calculated using periodic boundary conditions depends on the size of simulation box $L$ and tends to limit value of diffusion coefficient $D_0$ for infinite simulation box as
\begin{equation} \label{Dpbc}
D_{PBC}(L)=D_0 - \frac{A}{L}.
\end{equation}

To take this aspect into account for calculating the diffusion coefficient using both the VAF and MSD approaches, we provided simulations for each nanoparticle size in boxes of various sizes $L$ ranfing from 3 to 16\,nm. To derive diffusion coefficient $D_0$ we fitted the obtained dependencies of the diffusion coefficient on the simulation box size on the basis of eq. \eqref{Dpbc} varying $D_0$ and $A$ as parameters.

\begin{figure}
\includegraphics[width=1.0\linewidth]{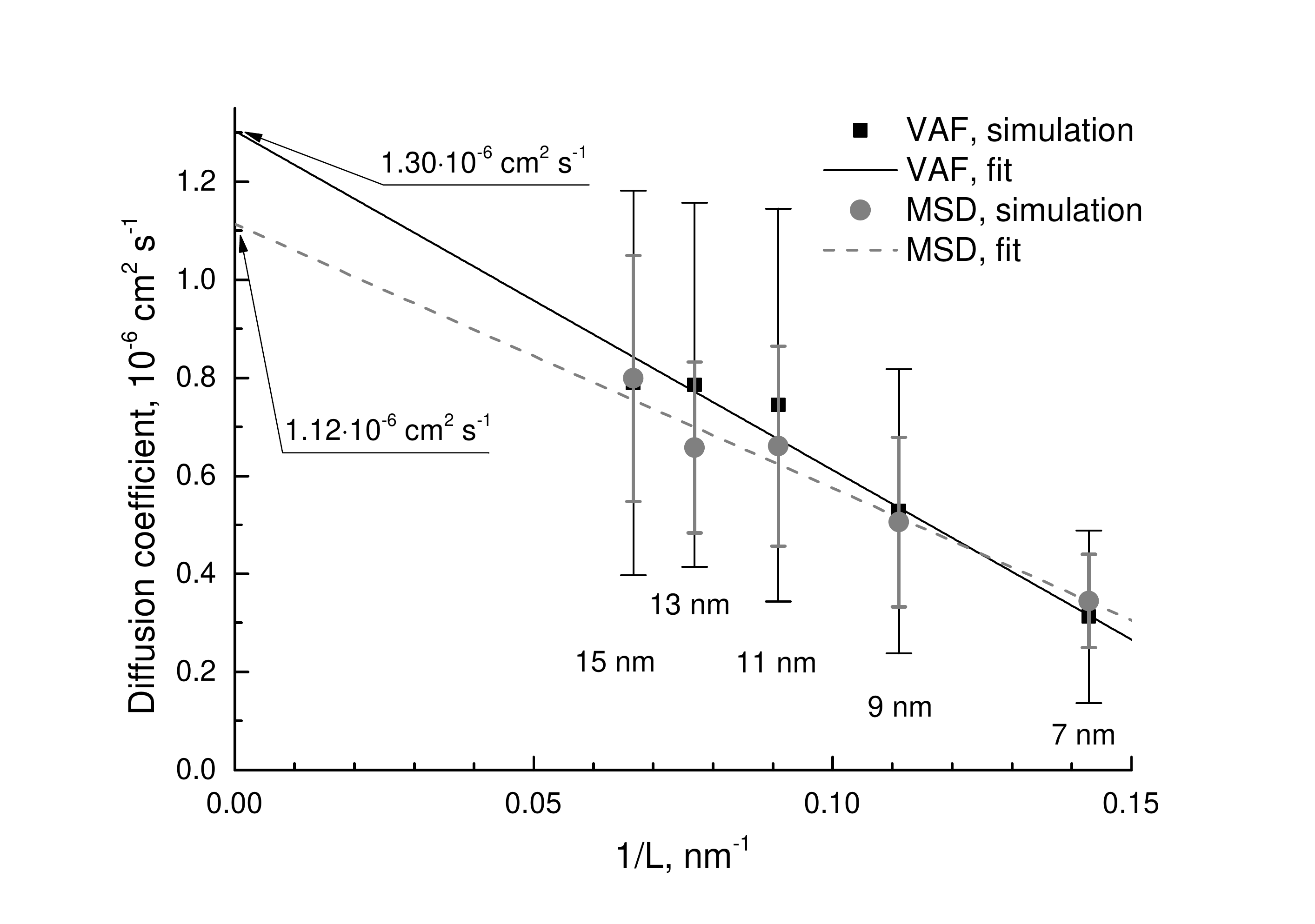}
\caption{Diffusion coefficient of 4\,nm particle calculated using the VAF (black squares) and MSD (grey circles) approach as a function of reciprocal simulation box size $1/L$ . The lines show the results of fit on the basis of eq. \eqref{Dpbc}. Treatment of the VAF data yields diffusion coefficient $D_0=1.30\cdot10^{-6}$\,cm$^2$s$^{-1}$ and treatment of the MSD data yields $D_0=1.12\cdot10^{-6}$\,cm$^2$s$^{-1}$. Error bars show standard deviations.}
\label{fig_4nm}
\end{figure}

Fig. \ref{fig_4nm} shows dependence of simulated diffusion coefficient for 4\,nm particle on the box size for the VAF and MSD approaches. The fit on the basis of eq. \eqref{Dpbc} yields $D_0=1.30\cdot10^{-6}$\,cm$^2$s$^{-1}$ for the VAF approach and $D_0=1.12\cdot10^{-6}$\,cm$^2$s$^{-1}$ for the MSD approach.

With decreasing the size of the solute, the definition of the radius figuring in the Stokes-Einstein relation becomes ambiguous \citep{BounadaryCond}. The radius can be interpreted as a bare radius of nanoparticle $R_{B}$ matching with the radius of crystallite or as hydrodynamic radius $R_{H} = R_{B}+\sigma$, where $\sigma \approx 0.145\,nm$ estimates the radius of the water molecule.

\begin{figure}
\includegraphics[width=1.0\linewidth]{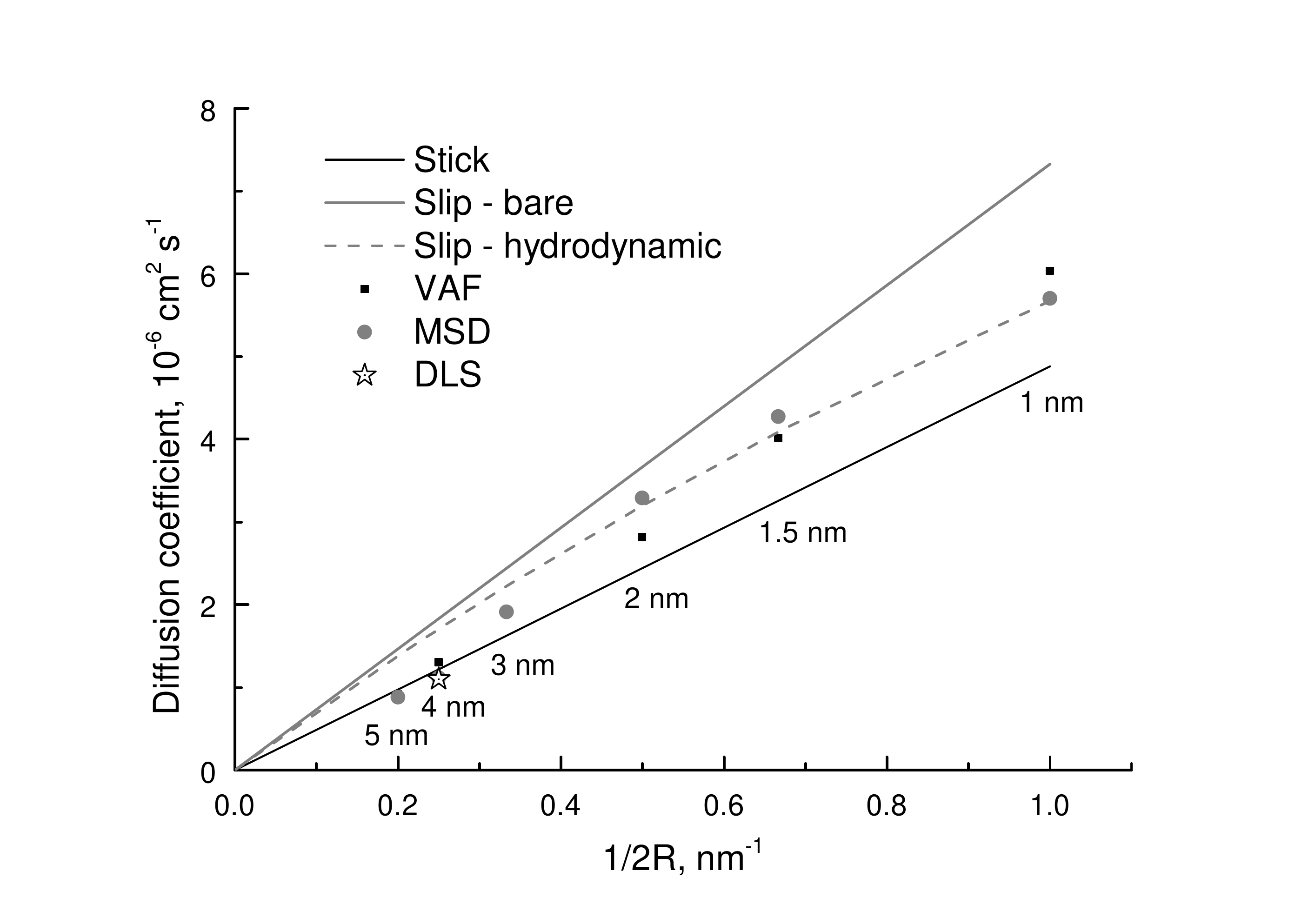}
\caption{Diffusion coefficients for particles of various diameters $2R$. The solid lines represent the predictions of Stokes-Einstein with bare radius used relation for stick (black) and slip (grey) boundary conditions. The dashed curve shows the Stokes-Einstein relation modified by the the hydrodynamic radius substitution. Black squares show the results of the VAF simulations and grey circles show the results of the MSD simulations. The star shows the value obtained from the DLS experiment for 4\,nm nanodiamonds.}
\label{fig_alltogehter}
\end{figure}

The dependencies of the diffusion coefficient on the particle size are shown in fig. \ref{fig_alltogehter} for different methods employed for their estimation. Table \ref{onecolumntable} lists corresponding numerical values of diffusion coefficients.

\begin{table*}[ht]
\caption{Diffusion coefficients for particles of various sizes. The predictions of the Stokes-Einstein relation (S-E) with stick and slip boundary conditions and the results of molecular dynamics simulations are listed. The diffusion coefficient of 4\,nm diamond particles obtained by DLS has an approximate value $1.1 \times 10^{-6} \, \text{cm}^2\text{s}^{-1}$.}
\begin{tabular}{lcccccr}
\hline
Approximate
& Number of
& Molecular
& \multicolumn{4}{l}{Diffusion Coefficient $D$, $10^{-6}$ cm$^2$ s $^{-1}$} \\
size, nm
& carbon atoms
& weight, kDa
&VAF
& MSD
&S-E, Stick
&S-E, Slip\\
\hline
1.0 & 87 & 1.05 & 6.03 & 5.70 &  4.88 & 7.33\\
1.5 & 293 & 3.52 & 4.02 & 4.26 & 3.25 & 4.88\\
2 & 729 & 8.76 & 2.82 & 3.29 & 2.44 & 3.66\\
3 & 2473 & 29.7 & 1.92 & 1.91 & 1.62 & 2.44\\
4 & 5851 & 70.3 & 1.30 & 1.12 & 1.22& 1.83\\
5 & 11543 & 138 & 0.91 & 0.88 &  0.98 & 1.47\\
\hline
\end{tabular}
\label{onecolumntable}
\end{table*}

\section{Conclusion}

We have considered three theoretical methods of nanoparticles diffusion coefficient estimation: the Stokes-Einstein relation with different type of hydrodynamic boundary conditions, the analysis of the molecular dynamics simulations employing the mean square displacement approach, and the analysis of the molecular dynamics simulations employing the Kubo-Green relation.

To calculate the diffusion coefficient via MD, it is strictly necessary to perform the series of simulations using the boxes of various sizes and taking the limit of an infinite box using formula \eqref{Dpbc}. The obtained effects of the finite box size are in agreement with predictions in \citep{PolymerChain,doi:10.1021/jp0477147}. It is noticeable that discrepancies between values calculated using VAF and MSD approaches are very small for all sizes of nanoparticles and boxes simulated, and one can not prefer one of these methods in this case.

The analysis of the diffusion coefficient allows to conclude that for particles larger than 3\,nm the Stokes-Einstein relation with stick boundary conditions describes the diffusion coefficient with high accuracy. For 2\,nm and smaller nanoparticles, the Stokes-Einstein relation with stick boundary conditions shows the tendency to underestimate the diffusion coefficient obtained in simulations. This is probably due to the transition of stick boundary conditions to slip boundary conditions with the decreasing particle size. The effect of hydrodynamic radius on the diffusion coefficients also becomes significant when the radius of the particle is smaller than 3 nm.

The principal result of the study is that the accuracy of the Stokes-Einstein relation is acceptable for the accurate measurements of hydrophobic nanoparticles size by dynamic light scattering. However the molecular dynamics simulation may be used for the refinement of the DLS data in the of small particles.

Contemporary facilities of the MD simulations allow varying the structure and the molecular, atomic, and ionic compound of the studied solute particles and the solvent. The proposed computational methods open a wide range of possibilities for investigating the diffusion parameters of nanoparticles in various types of solvents.

\begin{acknowledgments}
The authors express gratitude to A.Ya. Vul for his attention to this work. S.K. and A.S. acknowledge the project "Physical-chemical principles of new functionalized materials based on carbon nanostructures". E.E. acknowledges for support the Russian Foundation for Basic Research (project 12-08-00174a). Work was partly supported by the Skolkovo foundation and by the Russian Academy of Sciences. We thank the Joint Supercomputer Center of the Russian Academy of Sciences (JSCC RAS) for providing computational facilities.
\end{acknowledgments}

%%\bibliographystyle{spmpsci}
%\bibliographystyle{plainnat}

%\bibliography{Diffusion}

\begin{thebibliography}{25}%
\makeatletter
\providecommand \@ifxundefined [1]{%
 \@ifx{#1\undefined}
}%
\providecommand \@ifnum [1]{%
 \ifnum #1\expandafter \@firstoftwo
 \else \expandafter \@secondoftwo
 \fi
}%
\providecommand \@ifx [1]{%
 \ifx #1\expandafter \@firstoftwo
 \else \expandafter \@secondoftwo
 \fi
}%
\providecommand \natexlab [1]{#1}%
\providecommand \enquote  [1]{``#1''}%
\providecommand \bibnamefont  [1]{#1}%
\providecommand \bibfnamefont [1]{#1}%
\providecommand \citenamefont [1]{#1}%
\providecommand \href@noop [0]{\@secondoftwo}%
\providecommand \href [0]{\begingroup \@sanitize@url \@href}%
\providecommand \@href[1]{\@@startlink{#1}\@@href}%
\providecommand \@@href[1]{\endgroup#1\@@endlink}%
\providecommand \@sanitize@url [0]{\catcode `\\12\catcode `\$12\catcode
  `\&12\catcode `\#12\catcode `\^12\catcode `\_12\catcode `\%12\relax}%
\providecommand \@@startlink[1]{}%
\providecommand \@@endlink[0]{}%
\providecommand \url  [0]{\begingroup\@sanitize@url \@url }%
\providecommand \@url [1]{\endgroup\@href {#1}{\urlprefix }}%
\providecommand \urlprefix  [0]{URL }%
\providecommand \Eprint [0]{\href }%
\providecommand \doibase [0]{http://dx.doi.org/}%
\providecommand \selectlanguage [0]{\@gobble}%
\providecommand \bibinfo  [0]{\@secondoftwo}%
\providecommand \bibfield  [0]{\@secondoftwo}%
\providecommand \translation [1]{[#1]}%
\providecommand \BibitemOpen [0]{}%
\providecommand \bibitemStop [0]{}%
\providecommand \bibitemNoStop [0]{.\EOS\space}%
\providecommand \EOS [0]{\spacefactor3000\relax}%
\providecommand \BibitemShut  [1]{\csname bibitem#1\endcsname}%
\let\auto@bib@innerbib\@empty
%</preamble>
\bibitem [{\citenamefont {Li}(2009)}]{PhysRevE.80.061204}%
  \BibitemOpen
  \bibfield  {author} {\bibinfo {author} {\bibfnamefont {Z.}~\bibnamefont
  {Li}},\ }\href {\doibase 10.1103/PhysRevE.80.061204} {\bibfield  {journal}
  {\bibinfo  {journal} {Phys. Rev. E}\ }\textbf {\bibinfo {volume} {80}},\
  \bibinfo {pages} {061204} (\bibinfo {year} {2009})}\BibitemShut {NoStop}%
\bibitem [{\citenamefont {Tuteja}\ \emph {et~al.}(2007)\citenamefont {Tuteja},
  \citenamefont {Mackay}, \citenamefont {Narayanan}, \citenamefont {Asokan},\
  and\ \citenamefont {Wong}}]{TutejaEtAl2007}%
  \BibitemOpen
  \bibfield  {author} {\bibinfo {author} {\bibfnamefont {A.}~\bibnamefont
  {Tuteja}}, \bibinfo {author} {\bibfnamefont {M.~E.}\ \bibnamefont {Mackay}},
  \bibinfo {author} {\bibfnamefont {S.}~\bibnamefont {Narayanan}}, \bibinfo
  {author} {\bibfnamefont {S.}~\bibnamefont {Asokan}}, \ and\ \bibinfo {author}
  {\bibfnamefont {M.~S.}\ \bibnamefont {Wong}},\ }\href@noop {} {\bibfield
  {journal} {\bibinfo  {journal} {Nano Letters}\ }\textbf {\bibinfo {volume}
  {7}},\ \bibinfo {pages} {1276} (\bibinfo {year} {2007})}\BibitemShut
  {NoStop}%
\bibitem [{\citenamefont {Ould-Kaddour}\ and\ \citenamefont
  {Levesque}(2007)}]{ould-kaddour:154514}%
  \BibitemOpen
  \bibfield  {author} {\bibinfo {author} {\bibfnamefont {F.}~\bibnamefont
  {Ould-Kaddour}}\ and\ \bibinfo {author} {\bibfnamefont {D.}~\bibnamefont
  {Levesque}},\ }\href {\doibase 10.1063/1.2794753} {\bibfield  {journal}
  {\bibinfo  {journal} {The Journal of Chemical Physics}\ }\textbf {\bibinfo
  {volume} {127}},\ \bibinfo {eid} {154514} (\bibinfo {year}
  {2007})}\BibitemShut {NoStop}%
\bibitem [{\citenamefont {Schmidt}\ and\ \citenamefont
  {Skinner}(2003)}]{BounadaryCond}%
  \BibitemOpen
  \bibfield  {author} {\bibinfo {author} {\bibfnamefont {J.~R.}\ \bibnamefont
  {Schmidt}}\ and\ \bibinfo {author} {\bibfnamefont {J.~L.}\ \bibnamefont
  {Skinner}},\ }\href@noop {} {\bibfield  {journal} {\bibinfo  {journal} {The
  Journal of Chemical Physics}\ }\textbf {\bibinfo {volume} {119}} (\bibinfo
  {year} {2003})}\BibitemShut {NoStop}%
\bibitem [{\citenamefont {Johnson}\ and\ \citenamefont
  {Gabriel}(1994)}]{johnson1994lasers}%
  \BibitemOpen
  \bibfield  {author} {\bibinfo {author} {\bibfnamefont {C.}~\bibnamefont
  {Johnson}}\ and\ \bibinfo {author} {\bibfnamefont {D.}~\bibnamefont
  {Gabriel}},\ }\href {http://books.google.ru/books?id=B9-q3LV6xpkC} {\emph
  {\bibinfo {title} {Lasers Light Scattering}}},\ Dover Classics of Science and
  Mathematics Series\ (\bibinfo  {publisher} {Dover Publications,
  Incorporated},\ \bibinfo {year} {1994})\BibitemShut {NoStop}%
\bibitem [{\citenamefont {Pecora}(2000)}]{DLS_JNR}%
  \BibitemOpen
  \bibfield  {author} {\bibinfo {author} {\bibfnamefont {R.}~\bibnamefont
  {Pecora}},\ }\href {\doibase 10.1023/A:1010067107182} {\bibfield  {journal}
  {\bibinfo  {journal} {Journal of Nanoparticle Research}\ }\textbf {\bibinfo
  {volume} {2}},\ \bibinfo {pages} {123} (\bibinfo {year} {2000})}\BibitemShut
  {NoStop}%
\bibitem [{\citenamefont {Huang}\ \emph {et~al.}(2008)\citenamefont {Huang},
  \citenamefont {Sendner}, \citenamefont {Horinek}, \citenamefont {Netz},\ and\
  \citenamefont {Bocquet}}]{PhysRevLett.101.226101}%
  \BibitemOpen
  \bibfield  {author} {\bibinfo {author} {\bibfnamefont {D.~M.}\ \bibnamefont
  {Huang}}, \bibinfo {author} {\bibfnamefont {C.}~\bibnamefont {Sendner}},
  \bibinfo {author} {\bibfnamefont {D.}~\bibnamefont {Horinek}}, \bibinfo
  {author} {\bibfnamefont {R.~R.}\ \bibnamefont {Netz}}, \ and\ \bibinfo
  {author} {\bibfnamefont {L.}~\bibnamefont {Bocquet}},\ }\href {\doibase
  10.1103/PhysRevLett.101.226101} {\bibfield  {journal} {\bibinfo  {journal}
  {Phys. Rev. Lett.}\ }\textbf {\bibinfo {volume} {101}},\ \bibinfo {pages}
  {226101} (\bibinfo {year} {2008})}\BibitemShut {NoStop}%
\bibitem [{\citenamefont {Rudyak}\ \emph {et~al.}(2011)\citenamefont {Rudyak},
  \citenamefont {Krasnolutskii},\ and\ \citenamefont {Ivanov}}]{Rudyak}%
  \BibitemOpen
  \bibfield  {author} {\bibinfo {author} {\bibfnamefont {V.}~\bibnamefont
  {Rudyak}}, \bibinfo {author} {\bibfnamefont {S.}~\bibnamefont
  {Krasnolutskii}}, \ and\ \bibinfo {author} {\bibfnamefont {D.}~\bibnamefont
  {Ivanov}},\ }\href {\doibase 10.1007/s10404-011-0815-4} {\bibfield  {journal}
  {\bibinfo  {journal} {Microfluidics and Nanofluidics}\ }\textbf {\bibinfo
  {volume} {11}},\ \bibinfo {pages} {501} (\bibinfo {year} {2011})}\BibitemShut
  {NoStop}%
\bibitem [{\citenamefont {Korobov}\ \emph {et~al.}(2013)\citenamefont
  {Korobov}, \citenamefont {Volkov}, \citenamefont {Avramenko}, \citenamefont
  {Belyaeva}, \citenamefont {Semenyuk},\ and\ \citenamefont
  {Proskurnin}}]{C2NR33512C}%
  \BibitemOpen
  \bibfield  {author} {\bibinfo {author} {\bibfnamefont {M.~V.}\ \bibnamefont
  {Korobov}}, \bibinfo {author} {\bibfnamefont {D.~S.}\ \bibnamefont {Volkov}},
  \bibinfo {author} {\bibfnamefont {N.~V.}\ \bibnamefont {Avramenko}}, \bibinfo
  {author} {\bibfnamefont {L.~A.}\ \bibnamefont {Belyaeva}}, \bibinfo {author}
  {\bibfnamefont {P.~I.}\ \bibnamefont {Semenyuk}}, \ and\ \bibinfo {author}
  {\bibfnamefont {M.~A.}\ \bibnamefont {Proskurnin}},\ }\href {\doibase
  10.1039/C2NR33512C} {\bibfield  {journal} {\bibinfo  {journal} {Nanoscale}\
  }\textbf {\bibinfo {volume} {5}},\ \bibinfo {pages} {1529} (\bibinfo {year}
  {2013})}\BibitemShut {NoStop}%
\bibitem [{\citenamefont {Frenkel}(2013)}]{DarkSide}%
  \BibitemOpen
  \bibfield  {author} {\bibinfo {author} {\bibfnamefont {D.}~\bibnamefont
  {Frenkel}},\ }\href {\doibase 10.1140/epjp/i2013-13010-8} {\bibfield
  {journal} {\bibinfo  {journal} {The European Physical Journal Plus}\ }\textbf
  {\bibinfo {volume} {128}},\ \bibinfo {pages} {1} (\bibinfo {year}
  {2013})}\BibitemShut {NoStop}%
\bibitem [{\citenamefont {Baidakova}\ and\ \citenamefont
  {Vul'}(2007)}]{0022-3727-40-20-S14}%
  \BibitemOpen
  \bibfield  {author} {\bibinfo {author} {\bibfnamefont {M.}~\bibnamefont
  {Baidakova}}\ and\ \bibinfo {author} {\bibfnamefont {A.}~\bibnamefont
  {Vul'}},\ }\href {http://stacks.iop.org/0022-3727/40/i=20/a=S14} {\bibfield
  {journal} {\bibinfo  {journal} {Journal of Physics D: Applied Physics}\
  }\textbf {\bibinfo {volume} {40}},\ \bibinfo {pages} {6300} (\bibinfo {year}
  {2007})}\BibitemShut {NoStop}%
\bibitem [{\citenamefont {Ozerin}\ \emph {et~al.}(2008)\citenamefont {Ozerin},
  \citenamefont {Kurkin}, \citenamefont {Ozerina},\ and\ \citenamefont
  {Dolmatov}}]{Ozerin}%
  \BibitemOpen
  \bibfield  {author} {\bibinfo {author} {\bibfnamefont {A.}~\bibnamefont
  {Ozerin}}, \bibinfo {author} {\bibfnamefont {T.}~\bibnamefont {Kurkin}},
  \bibinfo {author} {\bibfnamefont {L.}~\bibnamefont {Ozerina}}, \ and\
  \bibinfo {author} {\bibfnamefont {V.}~\bibnamefont {Dolmatov}},\ }\href
  {\doibase 10.1134/S1063774508010070} {\bibfield  {journal} {\bibinfo
  {journal} {Crystallography Reports}\ }\textbf {\bibinfo {volume} {53}},\
  \bibinfo {pages} {60} (\bibinfo {year} {2008})}\BibitemShut {NoStop}%
\bibitem [{\citenamefont {Aleksenskii}\ \emph {et~al.}(1997)\citenamefont
  {Aleksenskii}, \citenamefont {Baidakova}, \citenamefont {Vul’},
  \citenamefont {Davydov},\ and\ \citenamefont
  {Pevtsova}}]{Raman_FTT_Pevtsova}%
  \BibitemOpen
  \bibfield  {author} {\bibinfo {author} {\bibfnamefont {A.}~\bibnamefont
  {Aleksenskii}}, \bibinfo {author} {\bibfnamefont {M.}~\bibnamefont
  {Baidakova}}, \bibinfo {author} {\bibfnamefont {A.}~\bibnamefont {Vul’}},
  \bibinfo {author} {\bibfnamefont {V.}~\bibnamefont {Davydov}}, \ and\
  \bibinfo {author} {\bibfnamefont {Y.}~\bibnamefont {Pevtsova}},\ }\href
  {\doibase 10.1134/1.1129989} {\bibfield  {journal} {\bibinfo  {journal}
  {Physics of the Solid State}\ }\textbf {\bibinfo {volume} {39}},\ \bibinfo
  {pages} {1007} (\bibinfo {year} {1997})}\BibitemShut {NoStop}%
\bibitem [{\citenamefont {Raty}\ and\ \citenamefont {Galli}(2003)}]{Raty}%
  \BibitemOpen
  \bibfield  {author} {\bibinfo {author} {\bibfnamefont {J.-Y.}\ \bibnamefont
  {Raty}}\ and\ \bibinfo {author} {\bibfnamefont {G.}~\bibnamefont {Galli}},\
  }\href {http://dx.doi.org/10.1038/nmat1018} {\bibfield  {journal} {\bibinfo
  {journal} {Nature Materials}\ }\textbf {\bibinfo {volume} {2}},\ \bibinfo
  {pages} {792} (\bibinfo {year} {2003})}\BibitemShut {NoStop}%
\bibitem [{\citenamefont {Aleksenskii}\ \emph {et~al.}(2012)\citenamefont
  {Aleksenskii}, \citenamefont {Shvidchenko},\ and\ \citenamefont
  {Eidel'man}}]{DLS_Shvid}%
  \BibitemOpen
  \bibfield  {author} {\bibinfo {author} {\bibfnamefont {A.}~\bibnamefont
  {Aleksenskii}}, \bibinfo {author} {\bibfnamefont {A.}~\bibnamefont
  {Shvidchenko}}, \ and\ \bibinfo {author} {\bibfnamefont {E.}~\bibnamefont
  {Eidel'man}},\ }\href {\doibase 10.1134/S1063785012120024} {\bibfield
  {journal} {\bibinfo  {journal} {Technical Physics Letters}\ }\textbf
  {\bibinfo {volume} {38}},\ \bibinfo {pages} {1049} (\bibinfo {year}
  {2012})}\BibitemShut {NoStop}%
\bibitem [{\citenamefont {Aleksenskiy}\ \emph {et~al.}(2011)\citenamefont
  {Aleksenskiy}, \citenamefont {Eydelman},\ and\ \citenamefont
  {Vul'}}]{Aleksenskiy}%
  \BibitemOpen
  \bibfield  {author} {\bibinfo {author} {\bibfnamefont {A.}~\bibnamefont
  {Aleksenskiy}}, \bibinfo {author} {\bibfnamefont {E.}~\bibnamefont
  {Eydelman}}, \ and\ \bibinfo {author} {\bibfnamefont {A.~Y.}\ \bibnamefont
  {Vul'}},\ }\href {\doibase doi:10.1166/nnl.2011.1122} {\bibfield  {journal}
  {\bibinfo  {journal} {Nanoscience and Nanotechnology Letters}\ }\textbf
  {\bibinfo {volume} {3}},\ \bibinfo {pages} {68} (\bibinfo {year}
  {2011})}\BibitemShut {NoStop}%
\bibitem [{\citenamefont {Konyakhin}\ \emph {et~al.}(2013)\citenamefont
  {Konyakhin}, \citenamefont {Sharonova},\ and\ \citenamefont
  {Eidelman}}]{Labeling_my}%
  \BibitemOpen
  \bibfield  {author} {\bibinfo {author} {\bibfnamefont {S.}~\bibnamefont
  {Konyakhin}}, \bibinfo {author} {\bibfnamefont {L.}~\bibnamefont
  {Sharonova}}, \ and\ \bibinfo {author} {\bibfnamefont {E.}~\bibnamefont
  {Eidelman}},\ }\href {\doibase 10.1134/S1063785013030073} {\bibfield
  {journal} {\bibinfo  {journal} {Technical Physics Letters}\ }\textbf
  {\bibinfo {volume} {39}},\ \bibinfo {pages} {244} (\bibinfo {year}
  {2013})}\BibitemShut {NoStop}%
\bibitem [{\citenamefont {Hess}\ \emph {et~al.}(2008)\citenamefont {Hess},
  \citenamefont {Kutzner}, \citenamefont {van~der Spoel},\ and\ \citenamefont
  {Lindahl}}]{doi:10.1021/ct700301q}%
  \BibitemOpen
  \bibfield  {author} {\bibinfo {author} {\bibfnamefont {B.}~\bibnamefont
  {Hess}}, \bibinfo {author} {\bibfnamefont {C.}~\bibnamefont {Kutzner}},
  \bibinfo {author} {\bibfnamefont {D.}~\bibnamefont {van~der Spoel}}, \ and\
  \bibinfo {author} {\bibfnamefont {E.}~\bibnamefont {Lindahl}},\ }\href
  {\doibase 10.1021/ct700301q} {\bibfield  {journal} {\bibinfo  {journal}
  {Journal of Chemical Theory and Computation}\ }\textbf {\bibinfo {volume}
  {4}},\ \bibinfo {pages} {435} (\bibinfo {year} {2008})},\ \Eprint
  {http://arxiv.org/abs/http://pubs.acs.org/doi/pdf/10.1021/ct700301q}
  {http://pubs.acs.org/doi/pdf/10.1021/ct700301q} \BibitemShut {NoStop}%
\bibitem [{\citenamefont {Mayo}\ \emph {et~al.}(1990)\citenamefont {Mayo},
  \citenamefont {Olafson},\ and\ \citenamefont
  {Goddard}}]{doi:10.1021/j100389a010}%
  \BibitemOpen
  \bibfield  {author} {\bibinfo {author} {\bibfnamefont {S.~L.}\ \bibnamefont
  {Mayo}}, \bibinfo {author} {\bibfnamefont {B.~D.}\ \bibnamefont {Olafson}}, \
  and\ \bibinfo {author} {\bibfnamefont {W.~A.}\ \bibnamefont {Goddard}},\
  }\href {\doibase 10.1021/j100389a010} {\bibfield  {journal} {\bibinfo
  {journal} {The Journal of Physical Chemistry}\ }\textbf {\bibinfo {volume}
  {94}},\ \bibinfo {pages} {8897} (\bibinfo {year} {1990})},\ \Eprint
  {http://arxiv.org/abs/http://pubs.acs.org/doi/pdf/10.1021/j100389a010}
  {http://pubs.acs.org/doi/pdf/10.1021/j100389a010} \BibitemShut {NoStop}%
\bibitem [{\citenamefont {Abascal}\ and\ \citenamefont {Vega}(2005)}]{water}%
  \BibitemOpen
  \bibfield  {author} {\bibinfo {author} {\bibfnamefont {J.~L.~F.}\
  \bibnamefont {Abascal}}\ and\ \bibinfo {author} {\bibfnamefont
  {C.}~\bibnamefont {Vega}},\ }\href {\doibase
  http://dx.doi.org/10.1063/1.2121687} {\bibfield  {journal} {\bibinfo
  {journal} {The Journal of Chemical Physics}\ }\textbf {\bibinfo {volume}
  {123}},\ \bibinfo {eid} {234505} (\bibinfo {year} {2005})}\BibitemShut
  {NoStop}%
\bibitem [{\citenamefont {Tazi}\ \emph {et~al.}(2012)\citenamefont {Tazi},
  \citenamefont {Boţan}, \citenamefont {Salanne}, \citenamefont {Marry},
  \citenamefont {Turq},\ and\ \citenamefont
  {Rotenberg}}]{0953-8984-24-28-284117}%
  \BibitemOpen
  \bibfield  {author} {\bibinfo {author} {\bibfnamefont {S.}~\bibnamefont
  {Tazi}}, \bibinfo {author} {\bibfnamefont {A.}~\bibnamefont {Boţan}},
  \bibinfo {author} {\bibfnamefont {M.}~\bibnamefont {Salanne}}, \bibinfo
  {author} {\bibfnamefont {V.}~\bibnamefont {Marry}}, \bibinfo {author}
  {\bibfnamefont {P.}~\bibnamefont {Turq}}, \ and\ \bibinfo {author}
  {\bibfnamefont {B.}~\bibnamefont {Rotenberg}},\ }\href
  {http://stacks.iop.org/0953-8984/24/i=28/a=284117} {\bibfield  {journal}
  {\bibinfo  {journal} {Journal of Physics: Condensed Matter}\ }\textbf
  {\bibinfo {volume} {24}},\ \bibinfo {pages} {284117} (\bibinfo {year}
  {2012})}\BibitemShut {NoStop}%
\bibitem [{\citenamefont {Markesteijn}\ \emph {et~al.}(2012)\citenamefont
  {Markesteijn}, \citenamefont {Hartkamp}, \citenamefont {Luding},\ and\
  \citenamefont {Westerweel}}]{viscosity}%
  \BibitemOpen
  \bibfield  {author} {\bibinfo {author} {\bibfnamefont {A.~P.}\ \bibnamefont
  {Markesteijn}}, \bibinfo {author} {\bibfnamefont {R.}~\bibnamefont
  {Hartkamp}}, \bibinfo {author} {\bibfnamefont {S.}~\bibnamefont {Luding}}, \
  and\ \bibinfo {author} {\bibfnamefont {J.}~\bibnamefont {Westerweel}},\
  }\href {\doibase http://dx.doi.org/10.1063/1.3697977} {\bibfield  {journal}
  {\bibinfo  {journal} {The Journal of Chemical Physics}\ }\textbf {\bibinfo
  {volume} {136}},\ \bibinfo {eid} {134104} (\bibinfo {year}
  {2012})}\BibitemShut {NoStop}%
\bibitem [{\citenamefont {Kubo}(1957)}]{JPSJ.12.570}%
  \BibitemOpen
  \bibfield  {author} {\bibinfo {author} {\bibfnamefont {R.}~\bibnamefont
  {Kubo}},\ }\href {\doibase 10.1143/JPSJ.12.570} {\bibfield  {journal}
  {\bibinfo  {journal} {Journal of the Physical Society of Japan}\ }\textbf
  {\bibinfo {volume} {12}},\ \bibinfo {pages} {570} (\bibinfo {year}
  {1957})}\BibitemShut {NoStop}%
\bibitem [{\citenamefont {Dunweg}\ and\ \citenamefont
  {Kremer}(1993)}]{PolymerChain}%
  \BibitemOpen
  \bibfield  {author} {\bibinfo {author} {\bibfnamefont {B.}~\bibnamefont
  {Dunweg}}\ and\ \bibinfo {author} {\bibfnamefont {K.}~\bibnamefont
  {Kremer}},\ }\href@noop {} {\bibfield  {journal} {\bibinfo  {journal} {The
  Journal of Chemical Physics}\ }\textbf {\bibinfo {volume} {99}} (\bibinfo
  {year} {1993})}\BibitemShut {NoStop}%
\bibitem [{\citenamefont {Yeh}\ and\ \citenamefont
  {Hummer}(2004)}]{doi:10.1021/jp0477147}%
  \BibitemOpen
  \bibfield  {author} {\bibinfo {author} {\bibfnamefont {I.-C.}\ \bibnamefont
  {Yeh}}\ and\ \bibinfo {author} {\bibfnamefont {G.}~\bibnamefont {Hummer}},\
  }\href {\doibase 10.1021/jp0477147} {\bibfield  {journal} {\bibinfo
  {journal} {The Journal of Physical Chemistry B}\ }\textbf {\bibinfo {volume}
  {108}},\ \bibinfo {pages} {15873} (\bibinfo {year} {2004})},\ \Eprint
  {http://arxiv.org/abs/http://pubs.acs.org/doi/pdf/10.1021/jp0477147}
  {http://pubs.acs.org/doi/pdf/10.1021/jp0477147} \BibitemShut {NoStop}%
\end{thebibliography}%


\end{document}